\documentclass[aps,twocolumn,preprintnumbers,amsmath,amssymb,longbibliography]{revtex4}
\usepackage{graphicx}% Include figure files
\usepackage{dcolumn}% Align table columns on decimal point
\usepackage{bm}% bold math
\usepackage{multirow}
\usepackage{color}
\usepackage{amsmath}
\usepackage{gensymb}
\usepackage[colorlinks=true]{hyperref}

\definecolor{mblue}{rgb}{0,0.35,0.75}
\newcommand{\addMisha}[1]{\textcolor{black}{#1}}

\begin{document}
\title{Charged excitons in monolayer WSe$_2$: experiment and theory}

\author{E. Courtade$^{1}$}
\author{M. Semina$^{2}$}
\author{M. Manca$^{1}$}
\author{M. M. Glazov $^{2}$}
\author{C. Robert$^{1}$}
\author{F. Cadiz$^{1}$}
\author{G. Wang$^{1}$}
\author{T. Taniguchi$^{3}$}
\author{K. Watanabe$^{3}$}
\author{M. Pierre$^4$}
\author{W. Escoffier$^4$}
\author{E. L. Ivchenko$^{2}$}
\author{P. Renucci$^{1}$}
\author{X. Marie$^{1}$}
\author{T. Amand$^{1}$}
\author{B. Urbaszek$^{1}$}

\affiliation{%
$^1$Universit\'e de Toulouse, INSA-CNRS-UPS, LPCNO, 135 Avenue de Rangueil, 31077 Toulouse, France}
\affiliation{$^2$Ioffe Institute, 194021 St.\,Petersburg, Russia}
\affiliation{$^3$National Institute for Materials Science, Tsukuba, Ibaraki 305-0044, Japan}
\affiliation{$^4$LNCMI-EMFL, INSA, UPS, UGA, CNRS-UPR3228, 143 Avenue de Rangueil, 31400 Toulouse, France}

\begin{abstract}
Charged excitons, or X$^{\pm}$-trions, in monolayer transition metal dichalcogenides have binding energies of several tens of meV. Together with the neutral exciton X$^0$ they dominate the emission spectrum at low and elevated temperatures. We use charge tunable devices based on WSe$_2$ monolayers encapsulated in hexagonal boron nitride, to investigate the difference in binding energy between X$^+$ and X$^-$ and the X$^-$ fine structure. We find in the charge neutral regime, the X$^0$ emission accompanied at lower energy by a strong peak close to the longitudinal optical (LO) phonon energy. This peak is absent in reflectivity measurements, where only the X$^0$ and an excited state of the X$^0$  are visible. In the $n$-doped regime, we find a closer correspondence between emission and reflectivity as the trion transition with a well-resolved fine-structure splitting of 6~meV for X$^-$ is observed. We present a symmetry analysis of the different X$^+$ and X$^-$ trion states and results of the binding energy calculations. We compare the trion binding energy for the $n$-and $p$-doped regimes with our model calculations for low carrier concentrations.  We demonstrate that the splitting between the X$^+$ and X$^-$ trions as well as the fine structure of the X$^-$ state can be related to the short-range Coulomb exchange interaction between the charge carriers.
\end{abstract}

%\pacs{78.60.Lc,78.66.Li} %Use showkeys class option if keyword

\maketitle

\section{Introduction}\label{sec:intro} 
The optical properties of transition metal dichalcogenides (TMDC) monolayers are dominated by excitons, electron-hole pairs bound by the attractive Coulomb interaction~\cite{He:2014a,Chernikov:2014a,Ugeda:2014a,Hanbicki:2015a,Wang:2015b,Qiu:2013a,Klots:2014a,Komsa:2012a,Ramasubramaniam:2012a,xuxuxu}. In the presence of additional charges, often due to non-intentional doping, also three particle complexes called trions (or charged excitons) can be observed, with binding energies of the order of 30~meV~\cite{Mak:2013a,Ross:2013a}. Trions in the solid state where originally reported for quantum wells at low temperature~\cite{Kheng:1993a}, and their existence is often associated to localisation effects. The first important difference for trions in TMDCs is that their their signature is not just observed at low temperature but up to room temperature \cite{Wang:2014b,McCreary:2016a,Shang:2015}. Other important differences compared to quantum well trions come from the very specific bandstructure of TMDC monolayers \cite{Kormanyos:2015a}:  The two non-equivalent valleys in momentum space can be addressed with chiral optical selections rules \cite{Xiao:2012a,Cao:2012a}, this allows to initialize the valley index. In addition, there exists a spin splitting in the conduction band (valence band) of several tens (hundreds) of meV \cite{Xiao:2012a,Kosmider:2013a,MolinaSanchez:2013,Zhu:2011a,Dery1}. This gives rise to many different valley and spin configurations between the three carriers, as for example in the negatively charged X$^-$ the extra electron can reside either in the same valley or in a different valley as compared to the photo-excited electron \cite{Yu:2014b}. \\
\indent In this work, we combine optical spectroscopy measurements with a theoretical analysis of the trion transitions. In order to observe spectrally narrow optical transition linewidth, that allow to study the fine-structure in detail, we encapsulate the WSe$_2$ monolayer in hexagonal boron nitride (hBN) \cite{Jin:2016a,Manca:2017a,Chow:2017a,Cadiz:2017a,Ajayi:2017a,Wang:2016v}. To switch electrically between the electron or hole-doped regimes, we have embedded the encapsulated monolayer in a charge tunable structure~\cite{Wang:2017a}. We observe the positively (X$^+$) and negatively (X$^-$) charged trions in reflectivity, with binding energies of about 20 and 30~meV, respectively. We measure a clear fine-structure splitting of the X$^-$ of 6~meV in both emission and absorption and we analyze the valley polarization of the fine structure components. Theoretical analysis is performed  to provide a symmetry classification of the trion states, which is rather intriguing as with the valley index an additional quantum number comes into play, going beyond the usual classification of trions in spin singlet and triplet states. We estimate trion binding energies of 20 to 30~meV for both X$^+$ and X$^-$ using an  effective mass approach. We demonstrate that for accepted values of effective masses the X$^+$ and X$^-$ binding energies should be identical, which is in contradiction to our experiments. We therefore argue that short-range Coulomb exchange effects provide reasonable X$^-$/X$^+$ splittings and result in X$^-$ fine structure \cite{Yu:2014b,Jones:2015a,Plechinger:2016b}.\\
\indent Here both experiment and theory are performed for low carrier concentrations where effects of screening are weak. A different, interesting prospect in \addMisha{TMD} monolayers is many body physics at high carrier densities \cite{suris:correlation,Efimkin:2017a,Dery:2016a,Back:2017a,Sidler:2016a} that can be probed in optics.\\
\indent \addMisha{The paper is organized as follows: In Sec.~\ref{sec:opto} we describe the sample and experimental results, in Sec.~\ref{sec:theory} the model and results of calculations are presented and in Sec.~\ref{sec:disc} the results are discussed and theory is compared with the experimental findings. The conclusion is given in Sec.~\ref{sec:concl}.} 

 \begin{figure*}
\includegraphics[width=0.8\textwidth]{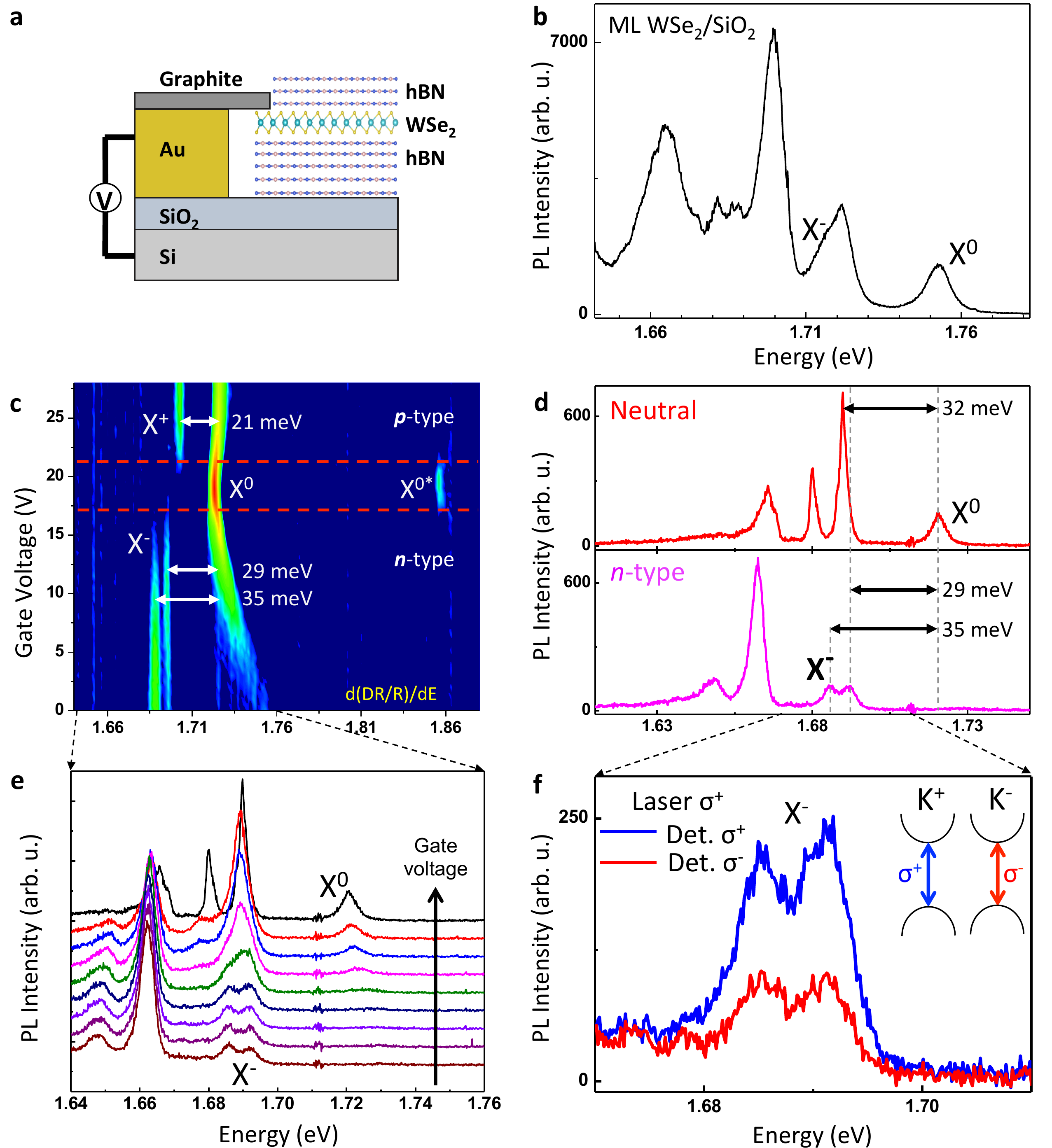}
\caption{\label{fig:fig1} (a) Schematics of the charge tunable van der Waals heterostructure. (b) Typical PL spectrum of an uncapped WSe$_2$ monolayer sample directly exfoliated onto SiO$_2$ is shown for comparison with our main results. The trion (X$^-$) and neutral exciton (X$^0$) peaks are indicated. (c) Contour plot of the first derivative \addMisha{with respect to energy} of the differential reflectivity. The $n$- and $p$-type regimes are clearly visible. (d) Typical PL response for the neutral regime (top) and the $n$-type regime (bottom). (e) Gradual evolution of the PL emission from the neutral to $n$-type regime. (f) Zoom on the trion PL transition X$^-$, detecting both circular polarization components following circularly polarized excitation. The inset shows a scheme of the chiral interband optical selection rules.
}
\end{figure*}

\section{Optical Spectroscopy}\label{sec:opto}

\subsection{Samples and Experimental Set-up}\label{sec:samples}
 The experiments are carried out at $T=4$~K  in a confocal microscope build in a vibration-free, closed cycle cryostat from Attocube. The excitation/detection spot diameter is $\approx1\mu$m, i.e. smaller than the typical ML diameter. The ML is excited by continuous wave He-Ne laser (1.96~eV) . The photoluminescence (PL) signal is dispersed in a spectrometer and detected with a Si-CCD camera. The white light source for reflectivity is a halogen lamp with a stabilized power supply. 
 
We have fabricated van der Waals heterostructures by mechanical exfoliation of bulk  WSe$_2$ (commercially available) and very high quality hBN crystals~\cite{Taniguchi:2007a}. \addMisha{A first layer of hBN was mechanically exfoliated and transferred onto a SiO$_2$ (90 nm)/Si substrate using PDMS stamping \cite{Gomez:2014a}. The deposition of the subsequent WSe2 ML and the second hBN capping layer was obtained by repeating this procedure to complete the full stack. We also transferred a thin graphite flake between the top surface of the WSe$_2$ ML and a Au pre-patterned electrode. Carrier concentration is varied by applying a bias between this electrode and the p-doped Si substrate (back gate).} The scheme of the structure is shown in Fig.~\ref{fig:fig1}a.
 
\subsection{Optical Spectroscopy Results}\label{sec:exper}
An overview of the neutral and charged exciton complexes in our sample is given in Fig.~\ref{fig:fig1}c. We measure differential reflectivity $(R_\text{ML}-R_\text{sub})/R_\text{sub}$, where $R_{\rm ML}$ is the intensity reflection coefficient of the sample with the WSe$_2$ monolayer and $R_{\rm sub}$ is the reflection coefficient of the hBN/SiO$_2$. We then plot its first derivative with respect to photon energy. 
Here our main target is to indicate the measured transition energies. To deduce quantitative information on the exact oscillator strength from reflectivity is very difficult due to possible interference effects in this van der Waals heterostructure \cite{Lien:2015}. But we can safely assume that transitions that are visible in this plot in Fig.~\ref{fig:fig1}c have a considerable oscillator strength for optical absorption \cite{Klingshirn:2007}. In the $p$-type region, where the Fermi level reaches the valence band, we observe the positively charged trion X$^+$ at an energy of 1.70~eV, i.e. 21~meV below the neutral exciton transition. In the neutral regime, where the semiconductor is non-degenerate as the Fermi level is between the valence and conduction band, we observe only the neutral exciton transition X$^0$ at 1.72~eV and an excited neutral exciton state X$^{0*}$, probably related to the $2s$ state of the A-exciton \cite{Chernikov:2014a,Wang:2015b}. In the $n$-doped regime, the negatively charged exciton X$^-$ appears, which shows a very clear fine structure splitting of 6~meV. The X$^-$ transitions are at 29 and 35~meV, respectively, below the X$^0$ transitions, so both energies are different from the LO phonon energy $E_{LO}=32$~meV~\cite{Zhang:2015a}.\\
\indent Now we compare these clear results in reflectivity/absorption, with the results obtained in photoluminescence emission. We plot two typical spectra in Fig.~\ref{fig:fig1}d. Surprisingly, we observe important differences: In the neutral regime, we observe the X$^0$ in PL at the same energy (within our error bars) as in reflectivity, indicating only a minor Stokes shift implying only weak localization of excitons. In addition to the X$^0$ (PL full width at half maximum, FWHM, down to $\approx4$~meV) we observe a very sharp peak at $32$ meV lower energy (FWHM down to $\approx1.6$~meV). This peak is totally absent in reflectivity, indicating a negligible oscillator strength. In terms of energy this peak is very close to the trion position. In contrast, in the $n$-type region, the neutral exciton PL emission disappears and the well defined double peak of the X$^-$ emerges, in agreement with the reflectivity results. Note that also the X$^-$ is accompanied by a sharp PL peak at lower energy that is not detected in reflectivity.\\
\indent In Fig.~\ref{fig:fig1}e we show how the PL emission gradually changes as we go from the neutral to the $n$-type regime. For intermediate bias values, see for example the pink and green curve, we have the trion and possibly phonon related emission superimposed, giving rise to a broader peak with two shoulders. This type of emission resembles the PL usually reported for the trion in ungated, uncapped structures, compare for instance with Fig.~\ref{fig:fig1}b, which shows a typical spectrum for a simple WSe$_2$ monolayer on SiO$_2$ \textit{not} encapsulated with hBN.\\ 
\indent It has been suggested that the X$^-$ fine structure splitting is induced by Coulomb exchange between the intravalley trion (both electrons in the same valley) and the intervalley trion (electrons in two different valleys) \cite{Yu:2014b}. First observations of trion PL emission with 2 components where interpreted accordingly \cite{Jones:2015a,Plechinger:2016b}.  In Fig.~\ref{fig:fig1}f we tried to find a difference in PL polarization between the two X$^-$ peaks, as suggested in \cite{Jones:2015a,Plechinger:2016b}. Here we excite with a $\sigma^+$ polarized laser, and the X$^-$ emission is strongly $\sigma^+$ polarized.  We do not find any noticeable difference between the high and low energy fine structure components. Different valley depolarization channels might in principle exist for each fine-structure peak if they correspond to intra- and intervalley \addMisha{trions, because, e.g., for the intravalley complex the spin-flip valley-conserving transition is forbidden, while this process may be allowed for the intervalley trion, see also Ref.~\cite{Singh}.} But we will observe no difference in stationary PL if the PL emission time is considerably shorter than the polarization decay time. Both times still need to be determined experimentally in hBN encapsulated samples.

\section{Theory}\label{sec:theory}

The main results from the experiments with high spectral resolution are a clear difference in X$^{+}$ and X$^{-}$ trion binding energies and a well-resolved fine structure splitting of the X$^{-}$ transition. In this section we estimate the trion binding energy, and we discuss why the X$^{+}$ and X$^{-}$ complexes have different binding energies and also the origin of the X$^{-}$ fine structure splitting. We give a symmetry analysis of the optically active and inactive trion states in monolayer WSe$_2$ that play a role optical spectroscopy experiments.

\addMisha{Below, in Sec.~\ref{sec:wave} we present the general approach to construct the three-particle wavefunction in the two-dimensional semiconductor and analyse the requirements imposed by the symmetry on the permutation of identical particles. Further, in Sec.~\ref{sec:binding} the effective Hamiltonian model for the envelope function of trions is introduced, the trial wavefunctions are presented and justified, and the trion binding energies are calculated. Then we move to the trion fine structure: Section~\ref{sec:sym} presents the results of the symmetry analysis of the X$^+$ and X$^-$ trion states and Sec.~\ref{sec:short} presents the model of the short-range exchange interaction in trions responsible for the trion states fine structure. }

\subsection{Trion wavefunctions}\label{sec:wave}
As a first step towards calculating the trion binding energies we need to define their wavefunction.
Owing to a sizable ($\gtrsim 100$~meV), spin-orbit splitting of the valence band it is sufficient to consider the hole states at the topmost valence band \cite{Cheiwchanchamnangij:2012a}, i.e., in a given valley we consider only one possible hole spin state. The hole Bloch state can therefore be labelled by a single quantum number ${\tau_v} = \pm 1$, denoting an unoccupied state at $\bm K_\pm$ valley at the edge of the Brillouin zone. This corresponds to the electron representation, where the Bloch function of the hole is $\mathcal U^h_{{\tau_v}}(\bm r_h) = \hat{\mathcal K} \mathcal U^{v.b.}_{-{\tau_v}} (\bm r_h)$, with $\mathcal U^{v.b.}_{\tau} (\bm r)$ being the valence band Bloch function and $\hat{\mathcal K}$ is the time reversal operator~\cite{Ivchenko:2005a}. {The equation $\mathcal U^h_{{\tau_v}}(\bm r_h) = \hat{\mathcal K} \mathcal U^{v.b.}_{-{\tau_v}} (\bm r_h)$ means that, under the time inversion, the state in the valley $\bm K_\pm$ is transferred to the state $\bm K_\mp$. In particular, an empty state in the valley $\bm K_+$ is equivalent to the hole state in the valley $\bm K_-$.} The Bloch state of a conduction band electron, $\mathcal U^{c.b.}_{s\tau}(\bm r)$, is labeled by two quantum numbers $s=\pm 1/2$ and $\tau=\pm 1$, where $\tau$ enumerates the valley and $s$ distinguishes the spin states within the valley, being the spin projection onto the normal to the sample $z$. In what follows we present the position vector $\bm r = (\bm \rho,z)$, with $z$ being its normal components and $\bm \rho$ being the two-dimensional vector in the plane of the monolayer. 

Generally, the trion wavefunction can be written as 
\begin{equation}
\label{trion:wf:gen}
\Psi_{i,j;k}(\bm r_{i}, \bm r_{j}, \bm r_k) = \frac{e^{\mathrm i \bm K \bm R}}{\sqrt{S}} \varphi(\bm \rho_{i}, \bm \rho_{j}) 
 \mathcal U^{(2)}_{ij}(\bm r_{i},\bm r_{j}) \mathcal U^{(1)}_{k}(\bm r_k),
\end{equation}
where the subscripts $i$ and $j$ denote the two identical carriers, namely, two electrons $e_1$ and $e_2$ for the X$^-$-trion or two holes $h_1$ and $h_2$ for the X$^+$ trion, $k$ denotes the unpaired carrier. 
%\tlse{* Can we decide throughout the paper if we use i,j or 1,2 everywhere ?*} 
In Eq.~\eqref{trion:wf:gen} $\bm R = [m_i (\bm \rho_{i} + \bm \rho_{j})+m_k\bm \rho_k]/M$ is the trion center of mass in-plane coordinate, $\bm K$ is the wavevector of the center of mass translational motion, $S$ is the normalization area, $m_i$ ($m_k$) is the mass of one of the identical (unpaired) carriers, $M=2m_i+m_k$ is the total trion mass, $ \varphi(\bm \rho_{1}, \bm \rho_{2})$ is the envelope function describing the in-plane relative motion of the charge carriers in the trion with $\bm \rho_{1,2} = \bm \rho_{i,j} - \bm \rho_k$ being the relative in-plane coordinates,  and $\mathcal U^{(2)}_{ij}(\bm r_i, \bm r_j)$ [$\mathcal U^{(1)}_{k}(\bm r_k)$] are the two identical particles [unpaired particle] Bloch function. \addMisha{The form of the trion wavefunction~\eqref{trion:wf:gen} is general and is not restricted to any particular mass ratio of electrons and holes, it implies only that the trion as a whole is free to move in the monolayer plane, so that its envelope function can be recast as a function of the center of mass $\bm R$ and relative coordinates $\bm \rho_1$ and $\bm \rho_2$. The three-particle Bloch function is recast as a combination of products of the individual charge carriers wavefunctions because the binding energy of the trion is much smaller than the band gap.} The wavefunction Eq.~\eqref{trion:wf:gen} must be antisymmetric with respect to the permutation of two identical particles $i$ and $j$~\cite{ll3_eng}. In the representation~\eqref{trion:wf:gen} we disregard the antisymmetrization of the functions of the electron and hole~\cite{birpikus_eng,Glazov:2014a,Glazov:2015a}, the effects of \addMisha{the exchange} interaction are addressed below, see Sec.~\ref{sec:short} and \ref{sec:disc}.

In order to fulfil the antisymmetry requirement for the trion wavefunction Eq.~\eqref{trion:wf:gen} we recast the basis two-particle Bloch functions $\mathcal U^{(2)}_{ij}(\bm r_i, \bm r_j)$ either as an antisymmetric or symmetric combination of the single particle Bloch functions
\begin{equation}
\label{2e}
\mathcal U^{(2)}_{ij}(\bm r_{i},\bm r_{j}) = \frac{1}{\sqrt{2}}
\begin{cases}
\mathcal U_{i}(\bm r_{i})\mathcal U_{j}(\bm r_{j}) - \mathcal U_{i}(\bm r_{j})\mathcal U_{j}(\bm r_{i}),\\
\mathcal U_{i}(\bm r_{i})\mathcal U_{j}(\bm r_{j}) + \mathcal U_{i}(\bm r_{j})\mathcal U_{j}(\bm r_{i}).
\end{cases}
\end{equation}
Correspondingly, the envelope function $\varphi(\bm \rho_1,\bm \rho_2)$ describing the relative motion of the identical particles is symmetric with respect to the permutation $\bm \rho_1 \leftrightarrow \bm \rho_2$ for the Bloch function in the top line of Eq.~\eqref{2e} and it is antisymmetric for the Bloch function in the bottom line of Eq.~\eqref{2e}. Hereafter we denote the trions as \emph{symmetric} or \emph{antisymmetric} in accordance with the symmetry of the \emph{envelope} function $\varphi(\bm \rho_1,\bm \rho_2)$. As a result, for symmetric trions two identical carriers cannot occupy the same Bloch state i.e. spin and/or valley index must differ. In conventional III-VI and II-VI quantum wells the symmetric trions are also known as the (spin) singlet trions, while antisymmetric trions are denoted as triplet trions~\cite{PSSB:PSSB387,PhysRevB.71.201312}.    

\subsection{Calculation of the exciton and trion binding energies}\label{sec:binding}
Using the trion wavefunctions defined above, we can now calculate the binding energies.
The envelope functions $\varphi(\bm \rho_1,\bm \rho_2)$ are the eigenfunctions of the effective mass two-particle Hamiltonian,
\begin{multline}
\label{H:tr}
\mathcal H_{tr} = - \frac{\hbar^2}{2\mu}\left[\Delta_1 + \Delta_2 + \frac{2\sigma}{\sigma+1} \nabla_1 \nabla_2 \right] \\
+ V( \rho_1) + V( \rho_2) - V(|\bm \rho_1 - \bm \rho_2|),
\end{multline}
where $\Delta_l$ and $\nabla_l$ are the Laplacian and gradient operators acting on functions of \addMisha{relative motion} $\bm \rho_l$ \addMisha{($l=1,2$)},  $\mu = m_em_h/(m_e+m_h)$ is the reduced mass of the electron-hole pair,  $\sigma = m_i/m_k$ is the ratio of effective mass of one of the identical carrier to the effective mass of the non-identical one, i.e., $\sigma = m_e/m_h$ for the X$^-$ trion and $\sigma=m_h/m_e$ for the X$^+$ one\addMisha{. Equation~\eqref{H:tr} is written in terms of the relative motion coordinates $\bm \rho_1$ and $\bm \rho_2$ of identical carriers with respect to the unpaired one, the term $\propto \nabla_1 \nabla_2$ accounts for a finite mass ratio $\sigma$ and known as Hughes-Eckart term in the theory of atoms and molecules. The kinetic energy $\hbar^2K^2/2M$ of the trion translational motion is excluded from Eq.~\eqref{H:tr}.}

\addMisha{In Eq.~\eqref{H:tr}} $V(\bm \rho)$ is the effective interaction potential taken in the form~\cite{Rytova67,1979JETPL..29..658K,Cudazzo:2011a,Chernikov:2014a,PhysRevB.88.045318,2017arXiv170107407S}: 
\begin{equation}
\label{Keldysh}
V(\rho) = -\frac{\pi }{2r_0 \varepsilon^*} \left[ \mathbf H_0\left( \frac{\rho}{r_0}\right) - Y_0 \left( \frac{\rho}{r_0}\right)\right],
\end{equation}
where $r_0$ is the effective screening radius, $\varepsilon^*$ is the effective dielectric constant being the average one of the dielectric constants of the substrate and cap layer, $\mathbf H_0$ and $Y_0$ are the Struve and Neumann functions. \addMisha{Note that in some works, e.g., in Ref.~\cite{Stier:2016a} the parameter $r_0$ is introduced in a different way with the factor $\varepsilon^*$ explicitly introduced in the arguments of the  $\mathbf H_0$ and $Y_0$ functions rather than in the prefactor of $V(\rho)$, namely, $V(\rho)= \pi/(2r_0)[\mathbf H_0(\rho\varepsilon^*/r_0) -Y_0(\rho\varepsilon^*/r_0)]$, this is simply equivalent to the rescaling $r_0 \to r_0/\varepsilon^*$.  } In Eq.~\eqref{Keldysh} we neglect a difference of interaction potentials of different charge carriers.
The difference, if any, is minor due to the atomic thickness of the TMD MLs. We assume that the screening parameters $r_0$ and $\varepsilon^*$ are the independent of frequency. We note that due to the significant binding energies of excitons, $\sim 10^2$~meV, and of trions, $\sim 10$~meV the screening of the Coulomb interaction in both cases may not be static, in general. Therefore we treat below $r_0$ and $\varepsilon^*$ as parameters of the theory, see Sec.~\ref{sec:disc} for discussion of particular values. Equations \eqref{H:tr} and \eqref{Keldysh} correspond to \textit{direct} electron-hole Coulomb interaction only, and in this Section~\ref{sec:wave} we disregard the short-range contributions to the electron-electron and the electron-hole interaction, discussed below in Sec.~\ref{sec:short}. \addMisha{We also neglect the possible lateral localization of trions in TMD ML plane extensively studied theoretically and experimentally in conventional semiconductor quantum well structures~\cite{filinov04,bracker05,Semina08e}. The in-plane localization can contribute to the inhomogeneous broadening of the trion lines in the spectra.}

The trion binding energy is the difference between the energies of the trion, i.e., the eigenenergy of the Hamiltonian~\eqref{H:tr}, and the energy of the neutral exciton~\footnote{The single-particle conduction band splittings do not contribute to the binding energy of the trion.}. The latter is found by minimizing the energy given by the effective exciton Hamiltonian in the form
\begin{equation}
\label{H:exc}
\mathcal H_{X} = - \frac{\hbar^2}{2\mu}\Delta + V(\rho),
\end{equation}
with $\bm \rho$ being the relative electron-hole coordinate. The exciton energy minimization is carried out using the (i)~the hydrogenic trial function
 \begin{subequations}
 \label{exciton}
\begin{equation}
\label{hydro}
\varphi_{ex}(\rho) \propto \exp{(-\alpha \rho)},
\end{equation}
with the single variational parameter $\alpha$, and (ii)~a more advanced trial function in the form
\begin{equation}
\label{hydro:1}
\varphi_{ex}(\rho) \propto \exp{(-\alpha \rho)} + \delta \rho\exp{(-\beta\rho)},
\end{equation}
\end{subequations}
 with two more parameters $\delta$ and $\beta$. The normalization constants are omitted in the trial functions. Equation~\eqref{hydro} has been used previously to calculate the binding energies of excitons in transition metal dichalcogenides monolayers~\cite{Chernikov:2014a,PhysRevB.88.045318}. We have also tested that the calculation with the advanced trial function gives the same binding energies as found by quantum Monte-Carlo calculations in Ref.~\cite{2017arXiv170107407S}. 

To calculate the binding energies of \emph{symmetric} trions over a whole range of the mass ratio $\sigma$ we used the sophisticated trial function suggested in Refs.~\cite{Sergeev:2001aa,Semina08e}
\begin{multline}
\label{trial:complex}
\varphi_s(\bm \rho_1, \bm \rho_2) \propto 
\\
\frac{\exp{(-\rho_1/a_1 - \rho_2/a_2)} + \exp{(-\rho_1/a_2 - \rho_2/a_1)}}{1 + d(|\bm \rho_1 - \bm \rho_2| - R_0){^2}}\times \\
\left(1+c|\bm \rho_1 - \bm \rho_2|\right){\exp\left(-s |\bm \rho_1 - \bm \rho_2| \right)},
\end{multline}
with the trial parameters $a_1$, $a_2$, $c$, $d$, $R_0$, and $s$. 
\addMisha{The choice of the trial function is motivated by the following: First, it contains the symmetrized combination of the exciton-like functions $\exp{(-\rho_1/a_1 - \rho_2/a_2)}$ for two carriers interacting with the unpaired one, the parameters $a_1$ and $a_2$ are the effective localization radii, such combinations can be viewed as wavefunctions for an exciton with another carrier bound to it. The factor $1+c|\bm \rho_1 - \bm \rho_2|$ accounts for the polarization of the complex and describes the repulsion of the paired carriers. This part of the wavefunction is known as Chandrasekar wavefunction used to describe H$^-$ ion with two light carriers bound to a heavier one~\cite{chandra1,chandra2}. Finally, the factors $\exp\left(-s |\bm \rho_1 - \bm \rho_2| \right)$ and $[1 + d(|\bm \rho_1 - \bm \rho_2| - R_0){^2}]^{-1}$ are included to describe the opposite limiting case of two heavy particles bound to a lighter one, i.e., the H$_2^+$-like case.} This function has been shown to produce high accuracy for conventional two-dimensional semiconductor systems based on III-V and II-VI quantum wells~\cite{Sergeev:2001aa}. We have compared the results of calculations using Eq.~\eqref{trial:complex} with the quantum Monte-Carlo results in Ref.~\cite{2017arXiv170107407S} and \addMisha{found good  accuracy of the suggested wavefunctions. For example, at $m_e=m_h$, $r_0/a_B=3/2$ we have $E_{tr}^b\approx0.06Ry$ and in Fig. 1 of Ref.~\cite{2017arXiv170107407S}  one has $E_{tr}^b \approx 0.075Ry$, at $r_0/a_B=1/4$ we have $E_{tr}^b=0.17Ry$ as compared with $0.2Ry$ in Ref.~\cite{2017arXiv170107407S}, at $r_0/a_B=1/9$ we have $E_{tr}^b \approx 0.24Ry$ and Ref.~\cite{2017arXiv170107407S} gives $0.26Ry$. For different masses, $m_e=2m_h$ and $r_0=a_B/4$ for the X$^-$ trion we obtain $0.19Ry$ as compared with $0.22Ry$ in Ref.~\cite{2017arXiv170107407S}. Here the dimensionless units corresponding to the exciton in a bulk system with the reduced mass $\mu$ and the dielectric constant $\varepsilon^*$ are introduced: the energy is measured in excitonic Rydbergs $Ry=\mu e^4/[{2}(\hbar\varepsilon^*)^2]$ and the length is measured in the excitonic Bohr radii  $a_B = \varepsilon^* \hbar^{{2}}/(\mu e^2)$.}

We have also calculated the binding energy of the \emph{antisymmetric} trion where the envelope function is antisymmetric with the replacement $\bm \rho_1 \leftrightarrow \bm \rho_2$. These are excited states and a reasonable trial function, being orthogonal to that in Eq.~\eqref{trial:complex}, takes the form~\cite{PSSB:PSSB387}
\begin{equation}
\label{triplet}
\varphi_{a}(\bm \rho_1, \bm \rho_2) \propto |\bm \rho_1 - \bm \rho_2| e^{\mathrm i \vartheta_{12}}\varphi_s(\bm \rho_1, \bm \rho_2),
\end{equation}
where $\vartheta_{12}$ is the angle of vector $\bm \rho_1 - \bm \rho_2$ with an in-plane axis and $\varphi_s(\bm \rho_1, \bm \rho_2)$ in introduced in Eq.~\eqref{trial:complex}. Again, the parameters of $\varphi_s(\bm \rho_1, \bm \rho_2)$, namely, $a_1$, $a_2$, $c$, $d$, $R_0$ and $s$ serve as  the variational parameters. 

 \begin{figure}
\includegraphics[width=\linewidth]{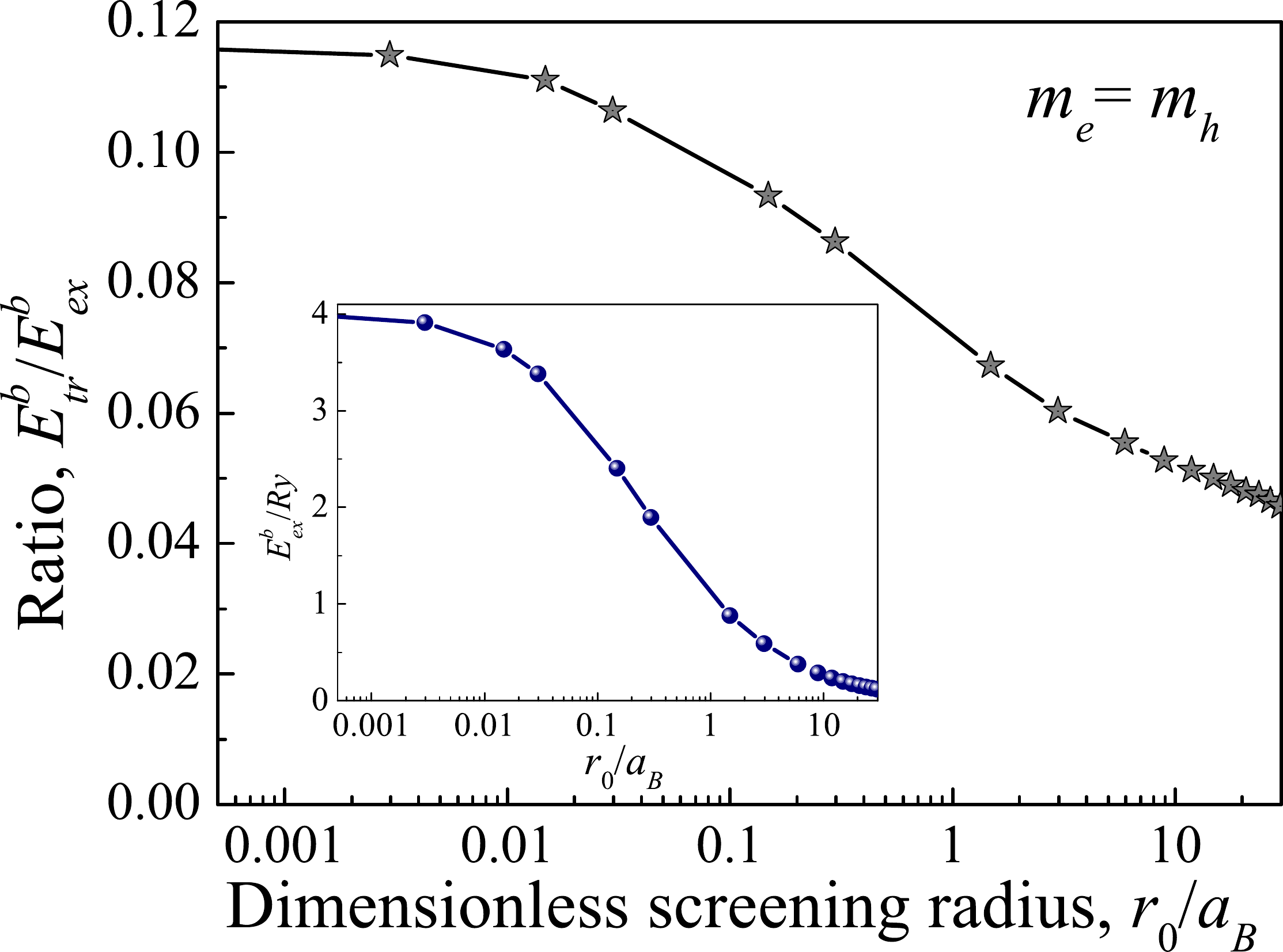}
\caption{Ratio of the trion binding energy $E_{tr}^b$ to the exciton binding energy $E_{ex}^b$ as a function of the screening radius $r_0$ at equal electron and hole effective masses. The inset shows the exciton binding energy vs. the screening radius. Units of energy and length are $Ry=\mu e^4/[\addMisha{2}(\hbar\varepsilon^*)^2]$ and $a_B = \varepsilon^* \hbar^{\addMisha{2}}/(\mu e^2)$, respectively.}\label{fig:rat} 
\end{figure}

The calculated ratio of trion, $E_{tr}^b$, and exciton, $E_{ex}^b$, binding energies as functions of the screening parameter $r_0$ for equal electron and hole effective masses is shown in Fig.~\ref{fig:rat}. The inset shows the exciton binding energy $E_{ex}^b$.  The screening radius $r_0=0$ corresponds to the strictly two-dimensional limit of a Coulomb problem where the exciton binding energy is $4~Ry$, while the trion binding energy is about $0.12~E_{ex}^b$~\cite{Stebe:1989aa,Thilagam:1997,Ivchenko:2005a,Sergeev:2001aa}. With an increase in $r_0$ the Coulomb potential becomes more shallow and both the exciton and trion binding energies decrease with $r_0$. For the same reason the ratio $E_{tr}^b/E_{ex}^b$ also decreases.

 \begin{figure}
\includegraphics[width=\linewidth]{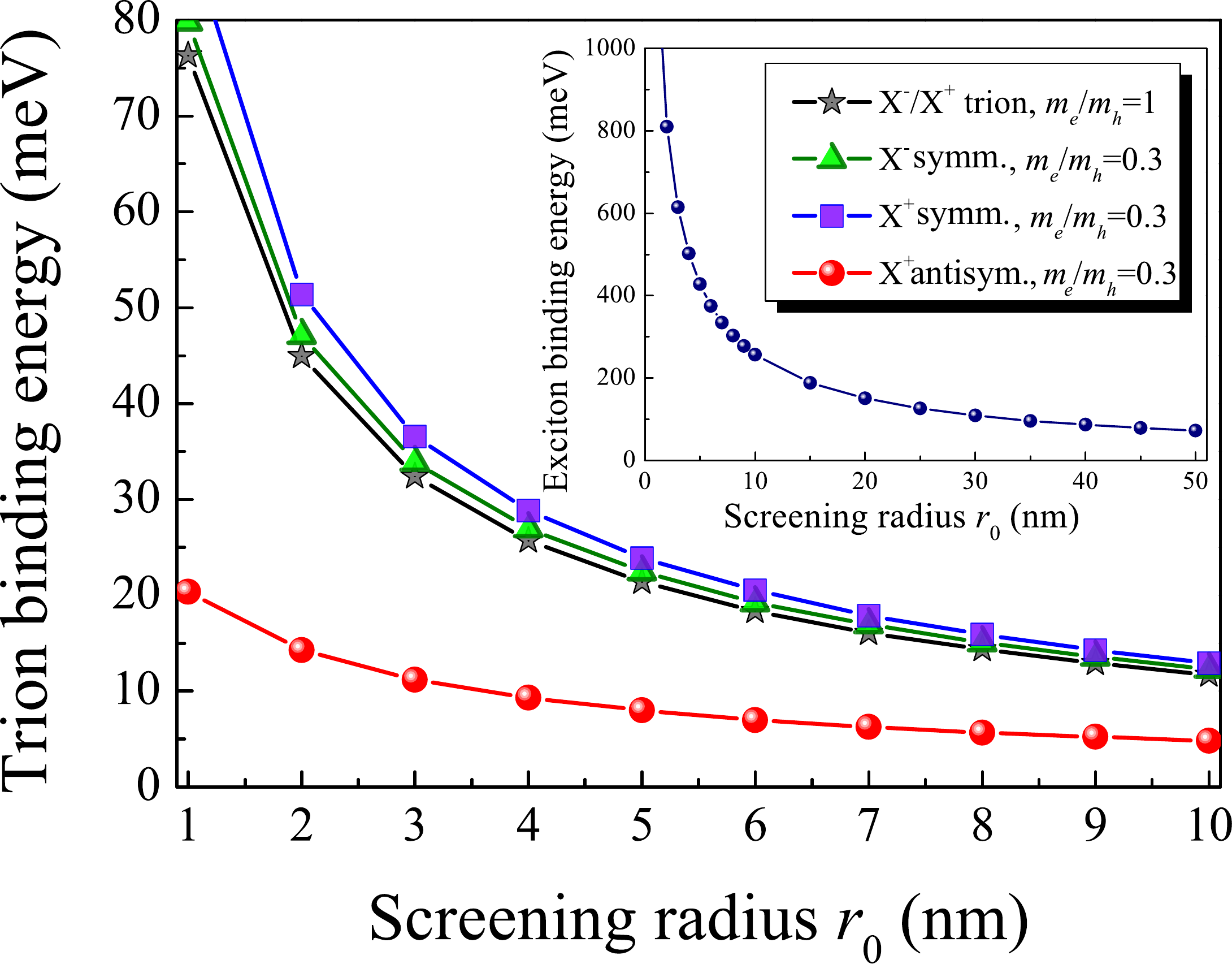}
\caption{Binding energies of the X$^+$ and X$^-$ trions for several effective mass ratio at a fixed reduced mass $\mu=0.16~m_0$, $\varepsilon^*=~1$. Inset shows the exciton binding energy as a function of the screening radius $r_0$. }\label{fig:r0} 
\end{figure}

Figure~\ref{fig:r0} demonstrates the results of calculation of the trion binding energies (main panel) and the exciton binding energy (inset) as a function of the screening radius $r_0$ in dimensional units. Here we took for simplicity  $\varepsilon^*=1$, the reduced mass $\mu=0.16~m_0$ with $m_0$ being the free electron mass, and considered two ratio of the effective masses $m_e/m_h=1$ and $m_e/m_h=0.3$. We obtain the exciton binding energies in the range of $\sim 10^2\ldots 10^3$~meV and the trion binding energies on the order of $10\ldots 100$~meV in agreement with previous calculations for exciton and symmetric trion binding energies~\cite{Chernikov:2014a,PhysRevB.88.045318,2017arXiv170107407S}. Note that the X$^{\pm}$ trion binding energies are not very sensitive to the effective mass ratio $m_e/m_h$. In Fig.~\ref{fig:r0} we also show the X$^+$-antisymmetric trion binding energy (red points) calculated for the electron-to-hole effective mass ratio $m_e/m_h=0.3$.

\begin{figure}
\includegraphics[width=\linewidth]{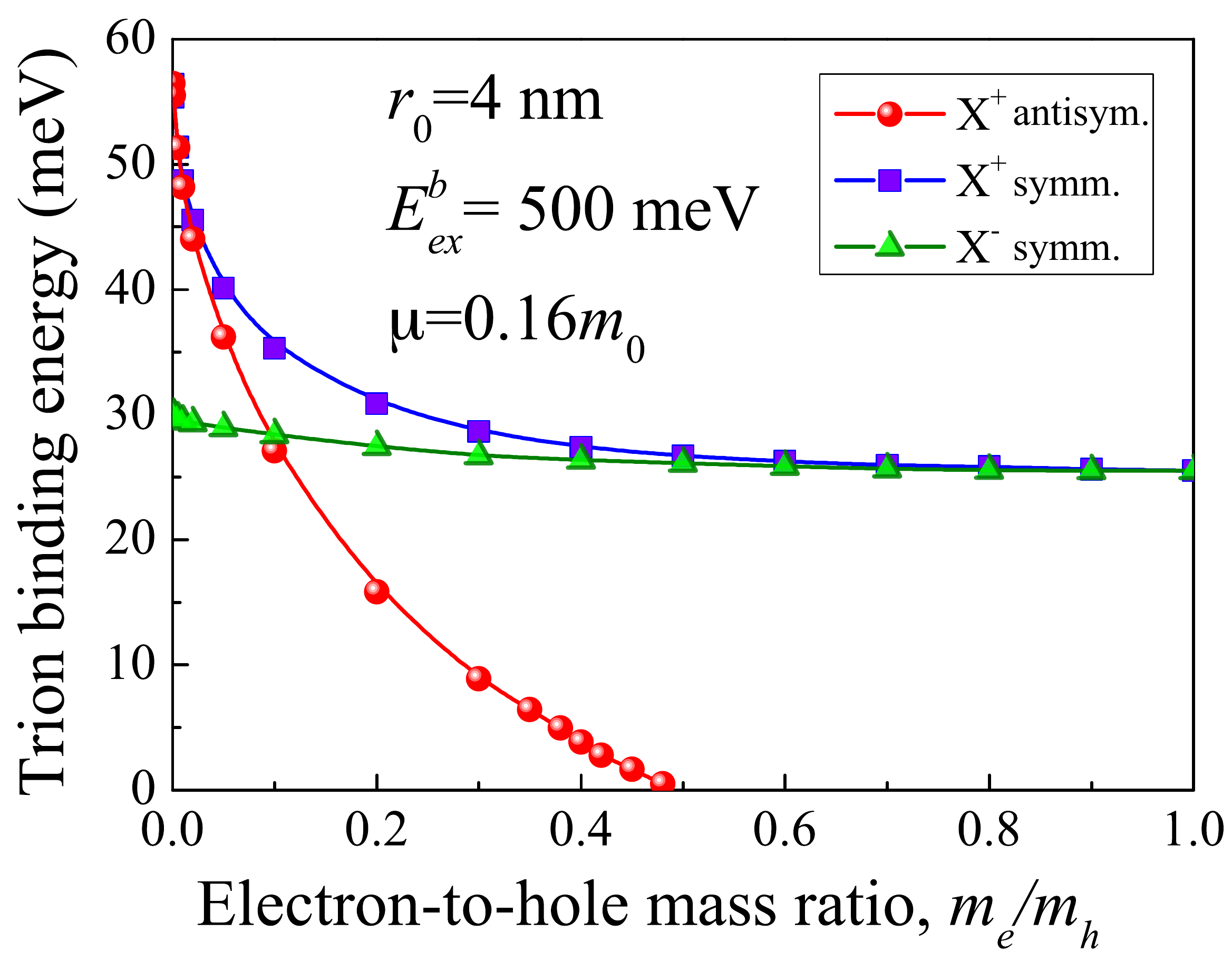}
\caption{Trion binding energies as a function of the effective mass ratio $\sigma$ at a fixed reduced mass $\mu=0.16~m_0$ and the screening radius $r_0=40$~\AA, $\varepsilon^*=1$. }\label{fig:sigma} 
\end{figure}

It is already seen from Fig.~\ref{fig:r0} that the trion with two heavier carriers, X$^+$ one in our case, has within the suggested model a higher binding energy. In order to study this effect in more detail we performed the calculations for fixed values of $\mu=0.16~m_0$ and $r_0=40$~\AA~which corresponds to the exciton binding energy for WSe$_2$ of 500~meV. The results of calculations are summarized in Fig.~\ref{fig:sigma}. At $\sigma = m_e/m_h\to 1$ the binding energies of X$^+$ and X$^-$ trions become equal, with the decrease in the mass ratio, $\sigma \to 0$,  the binding energies of trions increase. While the increase in the X$^-$ trion binding energy is quite minor, the increase in the X$^+$ trion binding energy is quite significant. Moreover, at a certain critical mass ratio $\sigma_{cr} \approx 0.5$ the antisymmetric X$^+$ trion state with the antisymmetric envelope function, Eq.~\eqref{triplet}, appears and becomes energetically stable. Its binding energy monotonously increases with a decrease in $m_e/m_h$. At small mass ratio $\sigma \lesssim 0.1$ the binding energy of the the antisymmetric state of X$^+$ trion exceeds that of the X$^-$ trion and approaches the binding energy of the symmetric X$^+$ trion at $\sigma\to 0$. 

\addMisha{We also note that for close values of the electron and hole masses $m_e\approx m_h$ one can use the simplified trial function where the parameters $d$, $c$, and $s$ are fixed to be zero~\cite{PhysRevB.88.045318}. Just like in quantum well structures with pure $1/r$ potential, it provides a reasonable accuracy of several percent to $10\ldots 20$\% due to a weak dependence of the trion binding energy on the mass ratio at $\sigma\to 1$, but gives the same values of X$^+$ and X$^-$ binding energies, see Sec.~\ref{sec:disc} for details. In contrast, the limit of $\sigma\to 0$ corresponds to the case of the donor-bound exciton. The wavefunction~\eqref{trial:complex} is similar to the ones used describe this situation~\cite{rashba,shi}.}

\subsection{Trion fine structure: symmetry analysis}\label{sec:sym}

As we so far included direct Coulomb terms only, the fact of having a valley index in addition to spin for each carrier did not impact our calculations. Analysing now the exact nature of the trion states will allow us to distinguish between optically active and inactive trions, that will contribute with their different recombination times to very complex emission and spin/valley polarization dynamics. Here and in what follows we consider the \emph{symmetric} trions only, because, as demonstrated above, the state with symmetric envelope is the ground state of the trion within the effective mass approximation. Moreover, the \emph{symmetric} trions are stable at arbitrary electron to hole mass ratio. 
To that end we apply group-theory analysis. Note that for the \emph{symmetric} trions where the envelope function $\varphi_s(\bm \rho_1, \bm \rho_2)$ is invariant under all transformations of the $D_{3h}$ point group, the trion wavefunction symmetry is given by the symmetry of the Bloch function, which transforms according to the representation 
\[
\mathcal D_{tr} = \mathcal D_i \times \mathcal D_j \times \mathcal D_k,
\]
where $\mathcal D_i$, $\mathcal D_j$, and $\mathcal D_k$ are the representations related with the Bloch functions, respectively, of two identical carriers, $i$ and $j$, and of the unpaired one, $k$.

\begin{table}[h]
\begin{center}
\caption{Symmetric X$^+$ trion states}
\begin{tabular}{c|cc|c|c}  
\hline
\multirow{2}{*}{\#} &\multicolumn{2}{c|}{State}  & \multirow{2}{*}{Representation of $D_{3h}$} & \multirow{2}{*}{} \\
 & $s_e$ & $\tau_e$ &   & \\
\hline
1 & $+1/2$ & $+1$ & \multirow{2}{*}{$\Gamma_9$} & $\sigma^+$\\
2 & $-1/2$ & $-1$ &  & $\sigma^-$\\
\hline
3 & $-1/2$ & $+1$ & \multirow{2}{*}{$\Gamma_8$} & \multirow{2}{*}{dark}\\
4 & $+1/2$ & $-1$ &  & \\
\hline
\end{tabular}\label{tab:tr1}
\end{center}
\end{table} 

\subsubsection{X$^+$ trion}\label{X+}

The X$^+$ trion is formed of two holes occupying the topmost valence band subbands and the unpaired electron. In the wavevector group $C_{3h}$ the valence band states transform according to the $\Gamma_7$ and $\Gamma_8$ irreducible representations in notations of Refs.~\cite{Glazov:2015a,Koster:1963a}. These two representations are compatible with $\Gamma_7$ representation of the $D_{3h}$ point group of the WSe$_2$ also relevant at the $\Gamma$ point. The product $\Gamma_7\times \Gamma_7= \Gamma_1+\Gamma_2 + \Gamma_5$ in $D_{3h}$ is reducible. The antisymmetric combination of the hole Bloch function in the top line in Eq.~\eqref{2e} forms spin and valley singlet and transforms according to the $\Gamma_1$ irreducible representation, i.e., it is invariant. The symmetry of the X$^+$ trion is, therefore, determined by the symmetry of the unpaired electron.

\begin{figure}
\includegraphics[width=\linewidth]{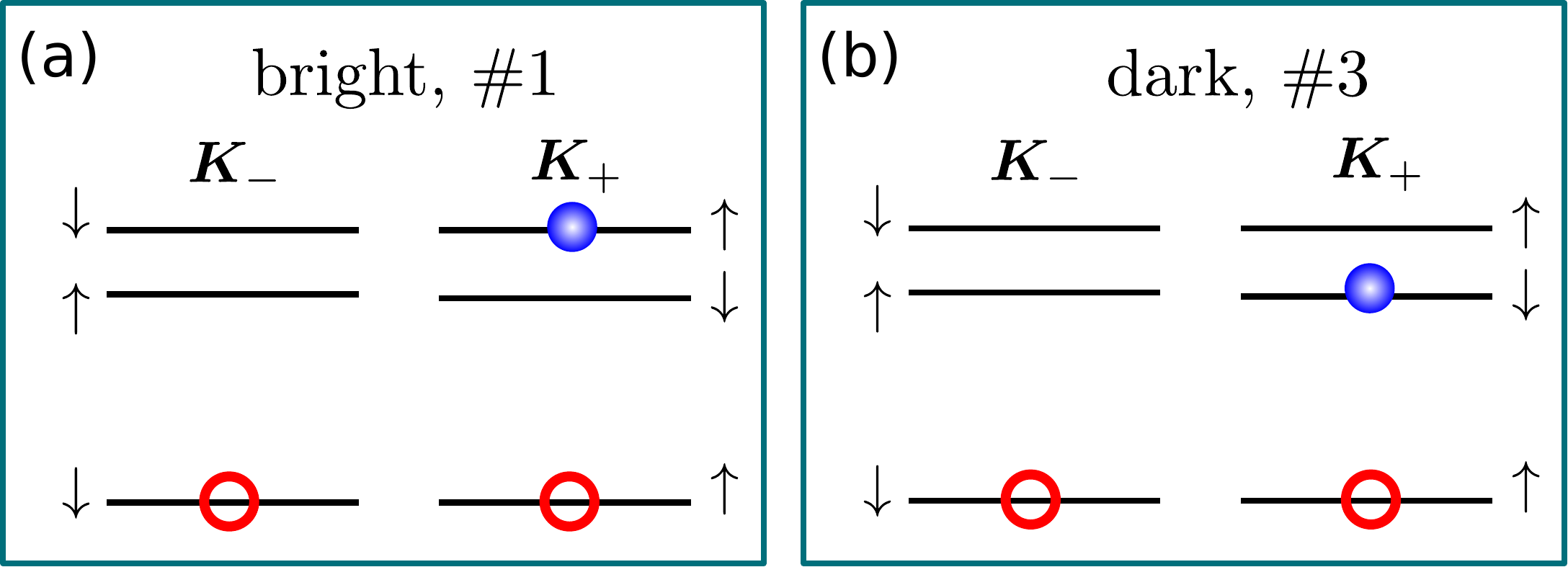}
\caption{Schematic illustration of the \emph{symmetric} X$^+$ trions: (a) state \#1 and (b) state \#3 in Tab.~\ref{tab:tr1}. Blue circles denote conduction band electron and open circles denote empty states in the valence band. The order of conduction subbands corresponds to WSe$_2$.}\label{fig:XP} 
\end{figure}

In WSe$_2$ the bottom conduction subbands and topmost valence subbands have opposite spins and the direct transitions at the normal incidence of radiation between these states are forbidden in the no-phonon processes. \addMisha{The transitions between the topmost valence subband and bottom conduction subband are possible in $z$-polarization within the same valley or with account for the electron-phonon interaction which changes the carriers valley. These processes studied in Refs.~\cite{dery2,basko,wang:z} and disregarded here.} In the wavevector group $C_{3h}$ the possible representations are $\Gamma_{11}$ and $\Gamma_{12}$, for the top subbands, where the optical transitions are possible, are compatible with the $\Gamma_9$ irreducible representation of the $D_{3h}$ point group. These two states form the bright doublet, states $1$ and $2$ in Tab.~\ref{tab:tr1}.  The two remaining states $3$ and $4$ in Tab.~\ref{tab:tr1} are formed with electrons in the bottom subbands of the conduction band, representations $\Gamma_{9}$ and $\Gamma_{10}$ of the $C_{3h}$ point group or $\Gamma_8$ of the $D_{3h}$ point group. These states are dark at normal light incidence in the no-phonon processes, because the direct interband transitions are forbidden between the topmost valence and bottom conduction subbands due to spin conservation law. The states $3,4$ can be activated in the phonon-assisted processes (involving, e.g., fully symmetric phonon, $A'$ in the wavevector group $C_{3h}$, with the wavevector $\bm K$ at the Brillouin zone edge) or due to the localization of the trions. In such a case the wavevector conservation law is relaxed and the processes where the electron changes the valley (but not spin) during the optical transition become possible. The examples of the bright and dark X$^+$ trion states are given in Fig.~\ref{fig:XP}.

Note that the optical selection rules are determined by both the symmetry of the initial state (valence band hole) and the final state (trion). At the normal incidence the components of the electric field transform according to the $\Gamma_6$ representation and, indeed, $\Gamma_6\times \Gamma_7 = \Gamma_8+\Gamma_9$. The presence of the $\Gamma_8$ representation demonstrates the possibility mentioned above to activate the dark X$^+$ trions in the $\Gamma_1$-phonon-assisted process.

\begin{figure}
\includegraphics[width=\linewidth]{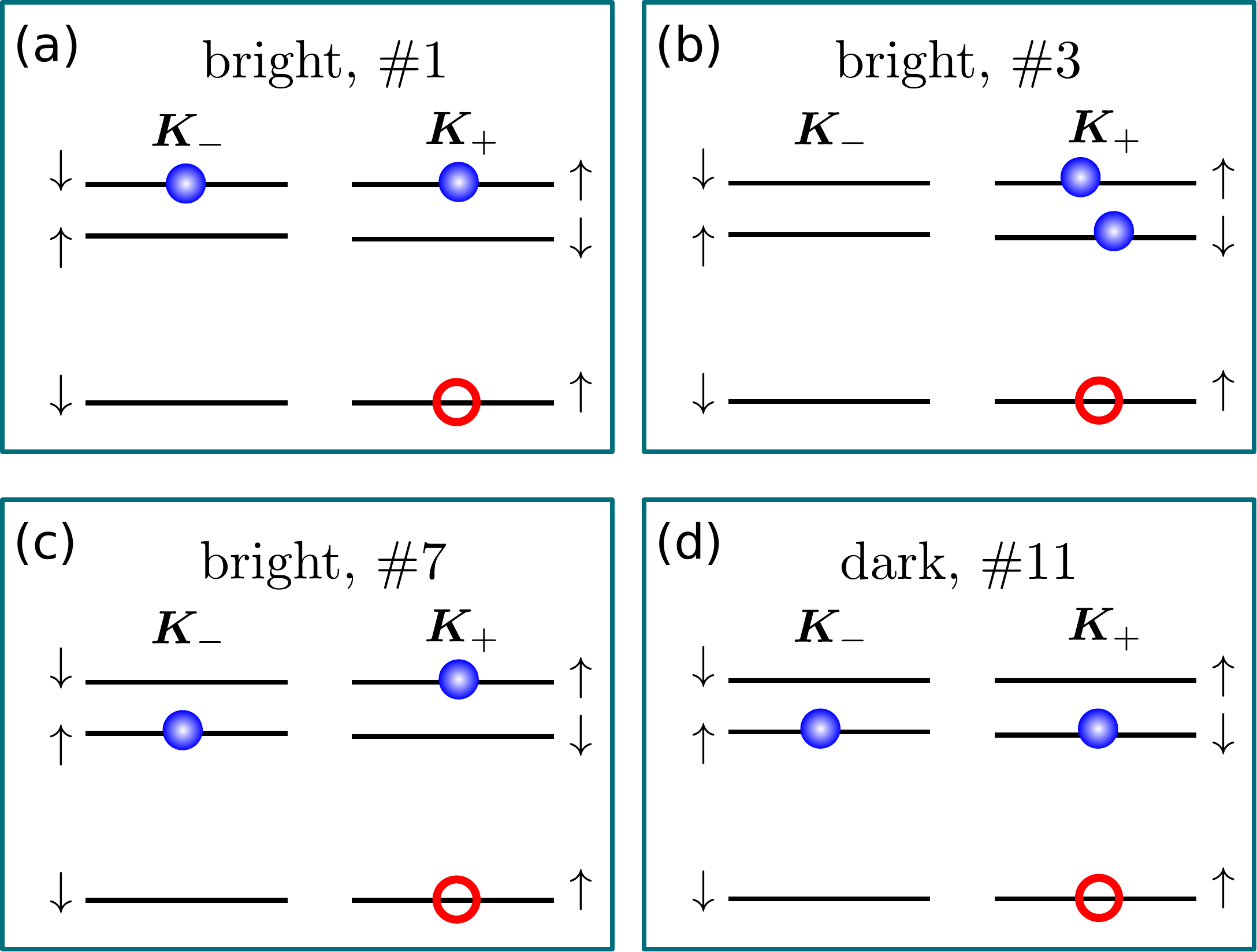}
\caption{Examples of the \emph{symmetric} X$^-$ trions: (a-c) optically active states, (d) dark state. Blue circles denote conduction band electron and open circles denote empty states in the valence band. The order of conduction subbands corresponds to WSe$_2$.}\label{fig:XM} 
\end{figure}

\subsubsection{X$^-$ trion}\label{X-} 
By contrast to the X$^+$ trions for the negatively charged trion a 12 \emph{symmetric} states are possible due to the moderate splitting between the conduction band spin states. Hence, the situation is more involved as compared with the X$^+$ case because there are six possible two-electron states using all spin and valley permutations. The relevant irreducible representations of the $D_{3h}$ point group can be found in a way described above using the following compatibility rules for the representations of $D_{3h}$ and $C_{3h}$ point groups:
\begin{align}
\Gamma_7^{{(D_{3h})}} \to \Gamma_7^{{(C_{3h})}}+\Gamma_8^{{(C_{3h})}},\\
  \Gamma_8^{{(D_{3h})}} \to \Gamma_9^{{(C_{3h})}}+\Gamma_{10}^{{(C_{3h})}},\\
  \Gamma_9^{{(D_{3h})}}\to \Gamma_{11}^{{(C_{3h})}}+\Gamma_{12}^{{(C_{3h})}}.
\end{align}
Here the left-hand side of equalities corresponds to $D_{3h}$, while the right-hand side corresponds to $C_{3h}$ point group. The bright and dark X$^-$ states are exemplified in Fig.~\ref{fig:XM}.  All 12 symmetric X$^-$ states are listed in Tab.~\ref{tab:tr2}. By contrast to Ref.~\cite{Danovich:2017aa} here we use the representations relevant for the point symmetry group of TMDC ML. For completeness, we also present in the table the irreducible representation of the two-electron Bloch function $\mathcal U_{s_1\tau_1,s_2\tau_2}^{(2)}$ corresponding to the top line of Eq.~\eqref{2e}. 

\begin{table}[h]
\begin{center}
\caption{Symmetric X$^-$ trion states. In parentheses the irreducible representations describing the transformation rule of the two-electron Bloch function are given. Superscript distinguishes equivalent representations of the two-electron Bloch function.  The irreducible representation of the hole state is $\Gamma_7$. For intervalley trions $\tau_1 \neq \tau_2$, for intravalley trions $\tau_1 = \tau_2$.}
\begin{tabular}{c|ccccc|c|c}  
\hline
\multirow{2}{*}{\#} &\multicolumn{5}{c|}{State}  & \multirow{2}{*}{Representation of $D_{3h}$} & \multirow{2}{*}{ } \\
 & $s_1$ & $\tau_1$ & $s_2$ & $\tau_2$ & ${\tau_v}$ &  & \\
\hline
1 & $+1/2$ & $+1$ & $-1/2$ & $-1$ & $+1$ & \multirow{2}{*}{$\Gamma_7$ ($\mathcal U_{ij}^{(2)}:~\Gamma_1^{{(1)}}$)} & $\sigma^+$\\ 
2 &$+1/2$ & $+1$ & $-1/2$ & $-1$ & $-1$ &  & $\sigma^-$\\ 
\hline
3 &$+1/2$ & $+1$ & $-1/2$ & $+1$ & $+1$ & \multirow{2}{*}{$\Gamma_9$ ($\mathcal U_{ij}^{(2)}:~\Gamma_6$)}  & $\sigma^+$\\ 
4 &$+1/2$ & $-1$ & $-1/2$ & $-1$ & $-1$ &  & $\sigma^-$\\ 
\hline
5 &$+1/2$ & $+1$ & $-1/2$ & $+1$ & $-1$ & \multirow{2}{*}{$\Gamma_8$ ($\mathcal U_{ij}^{(2)}:~\Gamma_6$)}  &  \multirow{2}{*}{dark} \\ 
6 &$+1/2$ & $-1$ & $-1/2$ & $-1$ & $+1$ &  &  \\ 
\hline
7 & $+1/2$ & $-1$ & $+1/2$ & $+1$ & $+1$ & \multirow{2}{*}{$\Gamma_7$ ($\mathcal U_{ij}^{(2)}:~\Gamma_5$)}  & $\sigma^+$\\ 
8 & $-1/2$ & $-1$ & $-1/2$ & $+1$ & $-1$ &  & $\sigma^-$\\ 
\hline
9 & $+1/2$ & $-1$ & $+1/2$ & $+1$ & $-1$ & \multirow{2}{*}{$\Gamma_9$ ($\mathcal U_{ij}^{(2)}:~\Gamma_5$)}  & \multirow{2}{*}{dark}\\ 
10 & $-1/2$ & $-1$ & $-1/2$ & $+1$ & $+1$ &  & \\ 
\hline
11 &$-1/2$ & $+1$ & $+1/2$ & $-1$ & $+1$ & \multirow{2}{*}{$\Gamma_7$ ($\mathcal U_{ij}^{(2)}:~\Gamma_1^{{(2)}}$)} &  \multirow{2}{*}{dark}\\ 
12 &$-1/2$ & $+1$ & $+1/2$ & $-1$ & $-1$ &  & \\ 
\hline
\end{tabular}\label{tab:tr2}
\end{center}
\end{table} 

\subsection{Short-range electron-electron exchange interaction}\label{sec:short}

The trion states listed in Tables I and II which transform according to the different irreducible representations of the $D_{3h}$ point group have, in general, different energies. The  states which transform according to the same irreducible representations, e.g., the X$^-$ states $(1,2)$, $(7,8)$, and $(11,12)$ or $(3,4)$ and $(9,10)$ can be mixed. In the effective mass model used above in Sec.~\ref{sec:binding} for binding energy calculations the envelope function $\varphi_s(\bm \rho_1,\bm \rho_2)$ is not sensitive to the trion Bloch function i.e. includes only the \textit{direct} Coulomb terms~\footnote{For generality, one can obtain the difference of the binding energies between the pair of states $(1,2)$, pair of states $(11,12)$, and remaining states $3\ldots 10$ taking into account that the effective masses of electrons in the ground and excited spin subbands in a given valley differ. The estimates show that this difference is minor and can be neglected.}. 
Hence, to understand splittings and possible mixing of  states the short-range electron-electron and electron-hole Coulomb \textit{exchange} interaction needs to be included in our analysis. 
The effective Hamiltonian of the short-range exchange interaction between the electron and the hole is a matrix in the space of spin/valley states of the electron-hole pair with the elements 
\begin{equation}
\label{eh:exch}
\hat V^{eh}(\bm \rho) = \delta(\bm \rho) \mathcal H_{\rm exch}^{eh},
\end{equation}
where the non-zero matrix elements $\langle s',\tau';{\tau_v}'| \mathcal H_{\rm exch}|s,\tau;{\tau_v}\rangle$ can be evaluated via the Bloch functions of the electron and hole~\cite{birpikus_eng,BP_exch71}. 

Similarly, the short-range part of the electron-electron interaction can be recast in the form:
\begin{equation}
\label{sr}
\hat{V}^{ee}(\bm \rho_1 - \bm \rho_2) = \delta(\bm \rho_1 - \bm \rho_2) \mathcal H_{\rm exch}^{ee},
\end{equation}
where the matrix elements  of the operator $\langle s_1',\tau_1';s_2',\tau_2'| \mathcal H_{\rm exch}|s_1,\tau_1;s_2,\tau_2\rangle$ can be expressed via the Bloch functions
\begin{multline}
\label{short-range}
\langle s_1',\tau_1';s_2',\tau_2'| \mathcal H_{\rm exch}^{ee}|s_1,\tau_1;s_2,\tau_2\rangle = \\
-\int d\bm r_{e1} d\bm r_{e2} U(\bm r_{e1} - \bm r_{e2})  \times \\ \left[ \mathcal U_{s_1'\tau_1'}(\bm r_{e2}) \mathcal U_{s_2'\tau_2'}(\bm r_{e1})\right]^*   \mathcal U_{s_1\tau_1}(\bm r_{e1}) \mathcal U_{s_2\tau_2}(\bm r_{e2}).
\end{multline}
Here $\mathcal U_{s\tau}(\bm r)$ is the electron Bloch function normalized per unit cell area $s_0$:
\begin{equation}
\label{bloch:norm}
\int_{v_0} d\bm r \left|\mathcal U_{s\tau}(\bm r)\right|^2=s_0,
\end{equation}
and $U(\bm r_{e1} - \bm r_{e2})$ is the  potential of the electron-electron interaction. It is noteworthy, that at small distances $|\bm r_{e1} - \bm r_{e2}| \sim a_0$, where $a_0$ is the lattice constant, the electron-electron interaction is strongly different from the effective potential~\eqref{Keldysh} and $U(\bm r_{e1} - \bm r_{e2})\propto e^2/|\bm r_{e1} - \bm r_{e2}|$ at $|\bm r_{e1} - \bm r_{e2}|\to 0$ because the screening is inefficient at atomic scales. Note that the details of the static screening of the short-range interaction in crystals are discussed in Refs.~\cite{PhysRevB.16.2717,zunger99,Voisin_exchange}. The integration in Eq.~\eqref{short-range} is carried out over the volume of the unit cell, so that $\bm r_{e1}$, $\bm r_{e2}$ are the three-dimensional position vectors.

Equation~\eqref{short-range} can be presented in the alternative form decomposing the products of the Bloch functions as~\cite{birpikus_eng}
\begin{multline}
\label{Bloch}
\left[\mathcal U_{s_2'\tau_2'}(\bm r)\right]^* \mathcal U_{s_1\tau_1}(\bm r) = \\ e^{\mathrm i (\bm K_{\tau_1} - \bm K_{\tau_2'}) \bm \rho} 
\sum_{M} B_M(z;s_2'\tau_2';s_1\tau_1)e^{-\mathrm i \bm b_M \bm \rho},
\end{multline}
where $\bm K_\tau$ is the wavevector of the valley $\tau=\pm 1$, $\bm b_M$ are the reciprocal lattice vectors and $B_M(z;s_2'\tau_2';s_1'\tau_1')$ are the coefficients, and introducing the Fourier components of the Coulomb interaction $U_{\bm q}(z) = \int d\bm \rho \exp{(\mathrm i \bm q \bm \rho)} U(r)$ with the result
\begin{multline}
\label{short-range1}
\langle s_1',\tau_1';s_2',\tau_2'| \mathcal H_{\rm exch}^{ee}|s_1,\tau_1;s_2,\tau_2\rangle
=\\
-  \sum_{L, M} \delta_{\bm q, \bm K_{\tau_1} - \bm K_{\tau_2'}-\bm b_M } \delta_{-\bm q, \bm K_{\tau_1'} - \bm K_{\tau_2}-\bm b_L }\times \\ \int dz_1 dz_2
B_M (z_2;s_2'\tau_2';s_1'\tau_1') \times \\
B_L (z_1;s_1\tau_1;s_2\tau_2) U_{\bm q}(z_1 - z_2).
\end{multline}

Two-electron states $|s_1,\tau_1;s_2\tau_2\rangle = \mathcal U^{(2)}_{s_1\tau_1;s_2\tau_2}(\bm r_1,\bm r_2)$ form a basis of the reducible representation which is decomposed into the irreducible representation $\Gamma_1^{(1)}, \Gamma_1^{(2)}, \Gamma_5, \Gamma_6$, see Tab.~\ref{tab:tr2}. Here the superscript $(1)$ or $(2)$ distinguishes equivalent irreducible representations relevant for the pairs $(1,2)$ and $(11,12)$. It is convenient to transform the matrix elements~\eqref{short-range1} from the basis $|s_1,\tau_1;s_2,\tau_2\rangle$ to the irreducible representations $\nu=\Gamma_1^{(1)}, \Gamma_1^{(2)}, \Gamma_5, \Gamma_6$. To establish the transformation rules for the two-electron Bloch functions from the basis $s_1\tau_1;s_2\tau_2$ to the basis $\nu=\Gamma_1^{(1)}, \Gamma_1^{(2)}, \Gamma_5, \Gamma_6$ we introduce the two sets of basic Pauli matrices $\bm \sigma^{(i)} = (\sigma_x^{(i)}, \sigma_y^{(i)},\sigma_z^{(i)})$ and $\bm \tau^{(i)} = (\tau_x^{(i)},\tau_y^{(i)},\tau_z^{(i)})$ acting in the spin and valley space of the $i$th ($i=1,2$) electron. Here the eigenstates of $\sigma_z$ operator with the eigenvalues $\pm 1$ correspond to the spin-up and spin-down electron and the eigenvalues $\tau_z=\pm 1$ of the corresponding valley operator correspond to the electron occupying the $\bm K_\pm$ valley, respectively. The expressions for the projection operators $\mathcal P_{\nu}$ to the trion states where the two-electron Bloch function transforms according to the irreducible representation $\nu$ can be recast as
\begin{align}
&\mathcal P_{\Gamma_5} = \frac{1-\bm \tau^{(1)} \cdot \bm \tau^{(2)}}{2} \frac{1+\sigma^{(1)}_z\sigma^{(2)}_z }{2}, \nonumber \\
&\mathcal P_{\Gamma_6} = \frac{1-\bm \sigma^{(1)} \cdot \bm \sigma^{(2)}}{2} \frac{1+\tau^{(1)}_z\tau^{(2)}_z }{2}, \nonumber \\
&\mathcal P_{\Gamma_1^{(1)}} = \frac{1}{\sqrt{2}}\left(\mathcal P_{\Gamma_1^{(S=0)}} + \mathcal P_{\Gamma_1^{(T=0)}} \right),  \\
&\mathcal P_{\Gamma_1^{(2)}} = \frac{1}{\sqrt{2}}\left(\mathcal P_{\Gamma_1^{(S=0)}} - \mathcal P_{\Gamma_1^{(T=0)}} \right),\nonumber
\end{align}
where
\begin{align}
\label{projections}
&\mathcal P_{\Gamma_1^{(S=0)}} = \frac{1-\bm \sigma^{(1)} \cdot \bm \sigma^{(2)}}{2} \frac{1 + \bm \tau^{(1)} \cdot \bm \tau^{(2)} - 2\tau^{(1)}_z\tau^{(2)}_z }{4}, \nonumber\\
&\mathcal P_{\Gamma_1^{(T=0)}} = \frac{1-\bm \tau^{(1)} \cdot \bm \tau^{(2)}}{2} \frac{1 + \bm \sigma^{(1)} \cdot \bm \sigma^{(2)} - 2\sigma^{(1)}_z\sigma^{(2)}_z }{4}, \nonumber.
\end{align} 
In this basis one has
\begin{equation}
\label{short-range2}
\mathcal H_{\rm exch}^{ee} 
=
a_0^2
\begin{pmatrix}
E_{\Gamma_1^{(1)}} & {V_{\Gamma_1}} & 0 & 0\\
{V_{\Gamma_1}} & E_{\Gamma_1^{(2)}} & 0 & 0\\
0 & 0 & E_{\Gamma_5} & 0 \\
0 & 0 & 0 & E_{\Gamma_6}
\end{pmatrix},
\end{equation}
where the parameters $E_\nu$, {$V_{\Gamma_1}$} have the dimension of energy and typically typically correspond to atomic energies, i.e., range from units to tens of eV. \addMisha{It immediately follows from the form of the exchange matrix elements in Eq.~\eqref{short-range1}, which is nothing but the matrix elements of the Coulomb interaction calculated over the Bloch functions within the unit cell. Accounting for the Bloch function normalization, Eq.~\eqref{bloch:norm}, one can crudely estimate $\langle s_1',\tau_1';s_2',\tau_2'| \mathcal H_{\rm exch}^{ee}|s_1,\tau_1;s_2,\tau_2\rangle$ as $\sim s_0 e^2/a_0$.} The splittings between the trion states are sensitive to the shape of the envelope function, cf. Eq.~\eqref{sr}, and are smaller, see Sec.~\ref{sec:disc} for discussion\addMisha{, because the exchange splitting is also proportional to the probability to find the electrons within the same unit cell}. Note that $V_{\Gamma_1}$ is non-zero because it mixes the states $\Gamma_1^{(1)}$ and $\Gamma_1^{(2)}$ of the same symmetry.

\section{Discussion}\label{sec:disc}

We now compare our experimental findings on trion binding energies in WSe$_2$ monolayers, Fig.~\ref{fig:fig1} with the model calculations presented in Sec.~\ref{sec:theory}. First of all, let us establish which of the X$^\pm$ trion states listed in Tabs.~\ref{tab:tr1} and \ref{tab:tr2} can manifest themselves in the reflectivity (absorption) or photoluminescence spectra.\\
\indent The situation is straightforward for the trions with two holes, X$^+$-trions. Here only two states, \#1 and 2, see Tab.~\ref{tab:tr1} and Fig.~\ref{fig:XP}(a), are optically active in, respectively, $\sigma^+$ and $\sigma^-$ circular polarizations at the normal incidence of radiation. Hence, in absence of magnetic field the X$^+$ trion produces a single line in absorption or PL spectra in agreement with experimental data shown in Fig.~\ref{fig:fig1}.\\
\indent Out of 12 symmetric X$^-$ trions three Kramers-degenerate pairs of trion states are optically active: $(1,2)$, $(3,4)$ and $(7,8)$. However, in WSe$_2$ the order of spin states in the conduction and valence band states is reversed~\cite{Kormanyos:2015a}, therefore the bright states $1$ and $2$ of the X$^-$ trion involve charge carriers from the topmost conduction subbands, Fig.~\ref{fig:XM}. The conduction band splitting is significant and amounts to about $37$~meV \cite{Kosmider:2013a}. As a result, for the reasonable electron densities, $n_e \lesssim 4\times 10^{12}$~cm$^{-2}$ and low temperatures of several Kelvin the occupancy of the excited subbands is negligible. Hence, the trion states \#1 and 2 are not active in absorption/reflection as they cannot be formed in the process of the single photon absorption \cite{12dark}.
Hence, in the conditions of our experiment only two pairs of states $(3,4)$ and $(7,8)$, Fig.~\ref{fig:XM}(b,c), are responsible for the two observed lines in the reflectivity, Fig.~\ref{fig:fig1}(c). Similarly, this doublet is seen in the PL spectra, Fig.~\ref{fig:fig1}(d), bottom panel.\\
\indent In accordance with our symmetry analysis the pairs of bright $(1,2)$ and dark $(11,12)$ states form two bases of the same $\Gamma_7$ irreducible representation and thus can be mixed by the parameter $V_{\Gamma_1}$ in Eq.~\eqref{short-range2}. This can result in a small but non-zero oscillator strength of  dark states $(11,12)$~\cite{Danovich:2017aa}. In accordance with our observations this mixing is negligible, because we do not observe third line related with X$^-$ trion neither in the reflectivity nor in PL, where, in principle, the small oscillator strength could be compensated by the significant occupancy of the trion state.\\
\indent Next, let us discuss the spectral positions of the observed X$^+$ and X$^-$ lines. Using the effective masses $m_e = 0.28~m_0$, $m_h = 0.36~m_0$~\cite{Kormanyos:2015a}, $\varepsilon^*=1$ and the screening parameter $r_0=40$~\AA~to reproduce the experimental exciton binding energy of about $E_{ex}^b=500$~meV \cite{He:2014a,Chernikov:2014a,Wang:2015b} we obtain almost equal binding energies of the X$^+$ and X$^-$ trions $E^b_{\mathrm X^-} \approx E^b_{\mathrm X^+} = (26\pm 1)$~meV, see Fig.~\ref{fig:sigma}. This value is in agreement with experimental data, Fig.~\ref{fig:fig1}(c), which correspond to slightly smaller X$^+$ binding energy of $21$~meV and slightly larger X$^-$ binding energy of $32$~meV measured from average position of two observe X$^-$ lines for very low $n$-type doping. Before addressing the difference of the positive and negative trion binding energies as well as the splitting of the X$^-$ doublet let us briefly analyze the role of dielectric environment described by the effective dielectric constant $\varepsilon^*$ on the trion binding energies. Our WSe$_2$ monolayer sample is encapsulated in hBN (see Fig.\ref{fig:fig1}a), whereas the results in particular for the exciton binding energy $E_{ex}^b$ measurements were obtained for monolayers directly in contact with the SiO$_2$/Si substrate. To investigate the influence of the dielectric environment we performed calculations using the parametrization of WSe$_2$ hBN heterostructure determined in Ref.~\cite{Stier:2016a} on the basis of analysis of excitonic diamagnetic shifts in this system: $\mu=0.18~m_0$, $r_0=13.6$~\AA, $\varepsilon^*=3.3$. The exciton binding energy is $E_{ex}^b=206$~meV in reasonable agreement with reported in Ref.~\cite{Stier:2016a} value of 221~meV, while the binding energy of X$^+$ and X$^-$ trions is $13$~meV, using $m_e=m_h$, which is somewhat lower than our experimental values show in Fig.\ref{fig:fig1}c. One can obtain the better with the experiment with the same value of $\varepsilon^*=3.3$ taking $\mu=0.16~m_0$, $r_0=6.4$~\AA~to obtain $E_{ex}^b=283$~meV and $E_{tr}^b=20.6$~meV.\\
\indent Our next aim is to analyze the role of the effective mass difference on the trion binding energies. \addMisha{Within the effective mass approximation, Eq.~\eqref{H:tr} the specific details of the bandstructure and Bloch states of individual carriers is reduced to the effective mass values $m_e$ and $m_h$.} Experimentally, we observe that the X$^+$ trion has a smaller binding energy as compared to X$^-$ in Fig.~\ref{fig:fig1}c. For the commonly used values $m_e<m_h$ from theory~\cite{Kormanyos:2015a} this is in contrast with the effective Hamiltonian calculations in Sec.~\ref{sec:binding} and Fig.~\ref{fig:sigma}. In the framework of this simple and practical model it is predicted that the trion with two heavier carriers has a larger binding energy. This is because the coefficient $2\sigma/(\sigma+1)$ in the ``correlation'' \addMisha{or  Hughes-Eckart} term in the kinetic energy, Eq.~\eqref{H:tr},
\begin{equation}
\label{crr}
-\frac{\hbar^2}{2\mu} \frac{2\sigma}{\sigma+1} \nabla_1\nabla_2
\end{equation}
 increases monotonously with an increase in $\sigma$. Thus, if one takes an optimal trial function of the system with lighter identical carriers and calculates the binding energy of the trion with heavier identical carriers this correlation term will produce a larger energy shift downwards. Hence, the binding energy of the trion with two heavier identical carriers is indeed expected to be larger. Since the effective conduction and valence band effective masses in WSe$_2$ are not precisely known, one may speculate that in this material $m_e>m_h$ and that is why the X$^-$ trion has a larger binding energy. However, to obtain the difference of about $10$~meV in the X$^+$ and X$^-$ binding energies one has to take an unrealistically large ratio $m_e>10m_h$. The difference of effective masses cannot also explain the splitting between two X$^-$ trions observed experimentally.

Hence, we resort to the assumption that $m_e$ and $m_h$ are close in magnitude in WSe$_2$ monolayers. Therefore, the difference of the X$^+$ and X$^-$ binding energies, as well as the splitting of the X$^-$ trion states, is provided by the short-range contributions to the exchange interaction analyzed in Sec.~\ref{sec:short}\addMisha{, which is particularly sensitive to the Bloch functions form}. In order to estimate these contributions we disregard the correlation term~\eqref{crr} in Eq.~\eqref{H:tr} and use the simplified trial function~\addMisha{\cite{chandra1,chandra2,PhysRevB.88.045318}}
\begin{multline}
\label{trial:simple}
\varphi_s(\bm \rho_1, \bm \rho_2) \propto\\
\left[\exp{(-\rho_1/a_1 - \rho_2/a_2)} + \exp{(-\rho_1/a_2 - \rho_2/a_1)}\right],
\end{multline}
with only two variational parameters $a_1$ and $a_2$. For the effective masses relevant for WSe$_2$ this trial function provides accuracy of several percents as compared to the more complex trial function\addMisha{, Eq.~\eqref{trial:complex}}. This is because, as shown in Fig.~\ref{fig:sigma} in the mass ratio range from $\sigma\approx0.6$ to $1$ the binding energies of the X$^+$ and X$^-$ practically merge. The evaluation of the short-range contribution to the trion energy after Eqs.~\eqref{sr} and \eqref{short-range2} using the trial function Eq.~\eqref{trial:simple} yields
\begin{equation}
\label{shift}
\delta E_{sr} = \frac{1}{2\pi} \frac{E_\nu a_0^2}{2\left(\frac{a_1a_2}{a_1+a_2}\right)^2 + \frac{(a_1+a_2)^2}{8}} = \frac{E_\nu a_0^2}{2\pi a_{\rm eff}^2}.
\end{equation}
Here
\[
a_{\rm eff} = \sqrt{2\left(\frac{a_1a_2}{a_1+a_2}\right)^2 + \frac{(a_1+a_2)^2}{8}}.
\]
For $a_0=3$~\AA~relevant for WSe$_2$, the estimates show that a reasonable difference of $|E_{\Gamma_5} - E_{\Gamma_6}|\approx 2$~eV\AA$^2$ is sufficient to produce the difference of the trion binding energies by $\approx 6$~meV in agreement with the fine structure splitting of the X$^-$ trion observed in the experiment. In a similar way, the short-range effects may produce the relative shift of the X$^+$ trion and the X$^-$ doublet in the optical spectra. The precise determination of parameters $E_\nu$ in Eq.~\eqref{short-range2} is beyond the scope of the present paper. Here we just stress that the short-range Coulomb exchange contributions to the trion energies and fine structure splittings provided by the exchange interaction give an order of magnitude of the X$^-$ fine structure and X$^+$ -- X$^-$-trions energy separation. \addMisha{Moreover, these estimates are consistent with the atomistic calculations of the bright and dark trion mixing matrix element, Eqs.~\eqref{short-range2}, \eqref{shift}, $\sim V_{\Gamma_1} a_0^2/(2\pi a_{\rm eff})\approx 20\ldots 30$~meV presented in Ref.~\cite{Danovich:2017aa}. } These contributions are expected to be particularly important in the two-dimensional transition metal dichalcogenides because of the small exciton and trion radii, as compared with the conventional semiconductor quantum wells.

\addMisha{In this work we focused on the case of WSe$_2$ monolayer. The theoretical analysis presented above is quite general and can be applied to other TMD material systems including both WS$_2$ and Mo-based monolayers. In the latter case of MoS$_2$ and MoSe$_2$ MLs the order of spin subbands in the conduction band is reversed. Hence, the X$^-$ state where both electrons occupy the bottom subband becomes optically active. This state is expected to dominate the reflectivity or absorption spectrum. Other active states where one electron occupies the excited spin subband either in the same or in the different valley can play a significant role in the reflectivity or absorption provided that the carrier density or temperature is high enough to produce the non-zero occupation of the excited spin subband.}

\section{Conclusion}\label{sec:concl}

In this work we have presented the results of experimental and theoretical study of the positively, X$^+$, and negatively, X$^-$, charged excitons in tungsten diselenide, WSe$_2$, monolayer. These Coulomb-correlated complexes comprising two holes and electron or two electrons and a hole are clearly observed in PL and reflectivity measurements performed in the van der Waals heterostructure based on the WSe$_2$ monolayer encapsulated in hexagonal boron nitride layers. The X$^+$ trion has a binding energy of $21$~meV, while the X$^-$ trion appears in the spectra as two peaks related with its energy spectrum fine structure at $29$~meV and $35$~meV below the exciton resonance.

The model describing the experimental findings is presented. Within the effective mass approach we evaluate the binding energies of the trions by means of the variational method using the trial functions which have previously proven to be reliable in conventional III-V and II-VI quantum well structures. We obtain the binding energies of the trions close to the experimentally observed values. We also provide the detailed symmetry analysis of the X$^+$ and X$^-$ trion states and identify the optically active and inactive configurations.  The fine-structure Hamiltonian for the X$^-$ trion is derived and the relation of its matrix elements with the Bloch functions is presented. We demonstrate that the fine structure of the observed X$^-$ emission as well as the splitting between the X$^+$ and X$^-$ trion is related with the short-range Coulomb exchange interaction between the charge carriers.

\acknowledgements 
We thank Reasmey Tan and Cherif Rouabhi at AIME for help with sample fabrication. We thank ERC Grant No. 306719, ITN Spin-NANO Marie Sklodowska-Curie grant agreement No 676108, ANR MoS2ValleyControl, Programme Investissements d'Avenir ANR-11-IDEX-0002-02, reference ANR-10- LABX-0037-NEXT, LIA ILNACS CNRS-Ioffe for financial support. X.M. also acknowledges the Institut Universitaire de France.  K.W. and T.T. acknowledge support from the Elemental Strategy Initiative
conducted by the MEXT, Japan and JSPS KAKENHI Grant Numbers JP26248061, JP15K21722
and JP25106006. M.M.G. and E.L.I. were supported by Russian Science Foundation Project No. 14-12-01067.

%%\bibliography{2dmatforTrion}% Produces the bibliography via BibTeX.

\begin{thebibliography}{65}
\bibitem[{12d()}]{12dark}
\bibinfo{note}{The X$^-$ states \#1 and 2 may be present in the
  photoluminescence spectra PL, however, the lifetime of these states and of a
  final single-electron state in the excited subband can be limited by the
  carrier relaxation towards the ground spin subband. This might produce weak
  and broad feature in the PL spectrum as compared with the PL of the states
  $(3,4)$ and $(7,8)$. So far, no indications of the states $(1,2)$ in the PL
  are observed.}
\end{thebibliography}


\end{document}